\documentclass{ws-p8-50x6-00}
\usepackage{amsmath}
\usepackage{graphicx}
\usepackage{eufrak}
\newlength{\dummysp}
\newcommand{\tr}{\mathop{{\hbox{Tr} \, }}\nolimits}

\newcommand{\stxt}[1]{\mathop{\hbox{{\scriptsize #1}}}\nolimits}
\newcommand{\bbar}[1]{{\overline{#1}}}

\newcommand{\beq}{\begin{eqnarray}}
\newcommand{\eeq}{\end{eqnarray}}

\newcommand{\e}{{\epsilon}}
\newcommand{\s}{{\sigma}}
\newcommand{\vev}[1]{{\langle #1 \rangle}}

\newcommand{\gappeq}{\mathrel{\rlap {\raise.5ex\hbox{$>$}}
{\lower.5ex\hbox{$\sim$}}}}
\newcommand{\lappeq}{\mathrel{\rlap{\raise.5ex\hbox{$<$}}
{\lower.5ex\hbox{$\sim$}}}}

\newcommand{\ite}[1]{\vspace{5pt} {\it #1 \hspace{2pt}}}

\newcommand{\ben}{\begin{enumerate}}
\newcommand{\een}{\end{enumerate}}

\newcommand{\sbar}{{\bar \s}}

\newcommand{\psib}{{\bar \psi}}

\newcommand{\bit}{\begin{itemize}}
\newcommand{\eit}{\end{itemize}}

\newcommand{\obf}{{\bf 1}}

\newcommand{\mbf}{{\bf m}}
\newcommand{\xb}{{\bbar{x}}}
\newcommand{\yb}{{\bbar{y}}}

\newcommand{\sbf}{\boldsymbol{\s}}

\newcommand{\Ncal}{{\cal N}}

\def\[{\left [}
\def\]{\right ]}
\def\({\left (}
\def\){\right )}

\begin{document}

\title{Lattice supersymmetry \\ and string phenomenology}

\author{Joel Giedt}

\address{Department of Physics, University of Toronto \\
60 Saint George Street, Toronto, ON M5S 1A7 Canada \\
E-mail:  giedt@physics.utoronto.ca}

\maketitle

\abstracts{I discuss the usefulness of lattice supersymmetry
in relation to string phenomenology.  I suggest how lattice results
might be incorporated into string phenomenology.  I outline difficulties
and describe some constructions that contain an exact lattice version of
supersymmetry, thereby reducing fine-tuning of the regulator.
I mention some problems that occur for these lattices.}

\ite{String phenomenology would benefit from lattice supersymmetry.}
Without stabilization of moduli, very little predictivity exists
in semi-realistic (supersymmetric) string-derived models.
The details of stabilization schemes are always
somewhat ``iffy'' because of nonholomorphic quantities
in the effective theory (things that depend on the
K\"ahler potential $K$, such as soft masses, B-terms, A-terms, and
more generally, the scalar potential), 
which are not protected by
nonrenormalization theorems.  Without such theorems,
one can only (reliably) resort to:
(i) symmetry constraints on the effective theory;
(ii) calculations in a weak coupling regime;
(iii) direct study of nonperturbative physics.
Typically, (i) and (ii) are not powerful enough
to yield the reliability that people like
myself long for.  On the other hand, (iii) may allow us to explore
some important issues in string phenomenology.  

Naturally, dynamical supersymmetry breaking 
through the physics of a strongly coupled 
gauge theory is a realm in which we would like to improve
our understanding of nonperturbative physics;
in particular, nonperturbative renormalization
of the K\"ahler potential.

But there are other aspects of the low energy effective theory that
depend on our understanding of strongly coupled gauge theories.
For example, it is well known that various {\it ad hoc}
assumptions for nonperturbative
corrections to the K\"ahler potential of the
dilaton can stabilize it at a weak coupling value;\cite{kahstab}
what is not as well appreciated is that {\it generic}
nonperturbative corrections to the K\"ahler potential of the
dilaton carve out local minima in the strong coupling
regime.\cite{Dine:1985he}  Until we understand the K\"ahler potential
in the strong coupling regime, we cannot really
say where the global minimum is.  Even if we fine-tune
beta function coefficients to yield a viable race-track
stabilization, a deeper minimum may exist at strong coupling
once the nonperturbative super-Yang-Mills (SYM) corrections
to the K\"ahler potential are included.  

The dilaton is just one of many moduli $\phi_i$ whose stabilization
we ultimately need to understand.  Thus a more general
statement of the observations made above is the following.
{\it Very often in semi-realistic string-derived models there
exist points in moduli space where at least one
factor in the gauge group becomes strongly coupled
in the infrared.  In such cases we often cannot reliably
say what the true vacuum of the theory is without an
improved understanding of the strongly coupled gauge theory.}
How can we say what are the deepest minima of the scalar
potential $V(K,W;\phi_i,\bar \phi_i)$ if we
only know $K$ in some subspace of the scalar manifold?
Lattice supersymmetry might be able to yield at least 
qualitative results related to this issue.

\ite{Nonperturbative super-QCD and supergravity.}
Consider how we typically incorporate super-QCD instantons
into an effective supergravity theory.
We obtain an effective action 
$S_{{\rm eff}} (g,\theta,\vev{\phi_i})$ 
that describes the instanton corrections from super-QCD,
which (unlike the supergravity we are supposed to be
studying) is a renormalizable field theory;
the $\vev{\phi_i}$ denotes that we often fix some
scalars to some point in the moduli space;
we also work at a fixed value of the gauge coupling $g$,
and the $\theta$ parameter dependence is generally implied
by holomorphy.  Then we
embed the instanton corrections back into the
supergravity effective action.  We argue that
it is okay to compute the instanton corrections
from super-QCD rather than the full supergravity
because the effects of the nonrenormalizable supergravity operators
in the effective theory will be supressed 
by the dynamical/fundamental hierarchy 
$\Lambda_{{\rm SQCD}} \ll m_P$.  There
is often a self-consistency criterion implicit in all of this; e.g.,
since $\Lambda_{{\rm SQCD}}$ depends on the dilaton,
taking $g,\theta$ to be frozen background
fields in the instanton calculation assumes that the
dilaton is sufficiently stabilized that we can
treat it this way; otherwise the results of the
computation with the dilaton held frozen may have
little to do with the actual instanton corrections
in the theory with a dynamical dilaton (particularly
where the K\"ahler potential is concerned).

By analogy, I would like to make the following ``lattice proposal'' for
studying nonperturbative corrections to string-derived
effective supergravity actions:
obtain $S_{{\rm eff}} (g,\theta,\vev{\phi_i})$ from lattice 
super-QCD.  Provided it is asymptotically free,
the {\it target continuum theory} 
has an ultraviolet attractive fixed
point, and therefore a corresponding lattice
theory should have a well-defined continuum limit.
The effective action will be obtained by the usual
procedure of matching lattice results for Euclidean Green's
functions to predictions of an effective (interpolating)
continuum action, up to errors of order $a \Lambda_{{\rm SQCD}}$,
where $a$ is the lattice spacing.  We then
embed these results into the ``tree'' supergravity
effective action, just as we do the instanton calculations.
In this way, the lack of {\it universality} associated
with a nonrenormalizable theory such as supergravity
is no big deal; it is just the usual regulator sensitivity
that one faces when computing quantum corrections
to an effective field theory.  What is important is that
our regulator respect the symmetries of the target
continuum theory.
The lattice action neglects nonrenormalizable supergravity
interactions; thus we expect to obtain results that
are valid up to order $1/a m_P$ corrections, as well
discretization and finite volume systematic errors.
What we need is a ``window'' where
$m_P \gg a^{-1} \gg \Lambda_{{\rm SQCD}}$.
In this way the lattice computation avoids the
physics of the underlying theory near the Planck scale.
Note that we will not be able to explore regions
of moduli space where this inequality fails to hold;
for this we need a nonperturbative analysis of
string theory itself.
We must also check the self-consistency of holding
moduli, such as the dilaton, fixed, just as we did
in the case of the instanton calculation.

\ite{Lattice supersymmetry is problematic.}
At most a discrete subgroup of the super-Poincar\'e
invariance group can be realized exactly on the lattice,
since it already breaks the Poincar\'e invariance
to a subgroup.  Moreover, it is well-known that
chiral symmetries are difficult to realize on
the lattice;\cite{Nielsen:1980rz} chiral R symmetries
and flavor symmetries play a key role in super-QCD.

``It's just a regulator,'' one might say.
True enough, in principle a combination of
(i) adding counterterm operators to the action, and
(ii) fine-tuning bare lattice parameters,
can yield the correct {\it quantum continuum limit:}
the continuum limit of lattice Green's functions 
involving energy-momentum scales well below the
cutoff should satisfy the various Ward identities.
However in practice fine-tuning and a complicated 
lattice action are ``expensive;'' the computations
take too long since the Green's functions must be
evaluated by Monte Carlo simulation over a range
of bare parameters in order to find a point that
will satisfy the Ward identities.  This can render the fine-tuning 
approach impractical.  Nevertheless quite a bit of 
work in this direction has been done; e.g., see 
refs.~\ref{Curci:1986sm}.

\ite{Exact lattice symmetries.}  For the hypercubic lattice
actions that are most often studied, discrete rotation
and translation symmetries guarantee Poincar\'e invariance
in the quantum continuum limit, without any fine-tuning:  
the exact lattice symmetries prevent one 
from writing down relevant or
marginal operators that would violate Poincar\'e invariance
in the {\it interpolating continuum action} (the effective action
that describes the quantum continuum limit of the lattice theory).  Long ago,
Ginsparg and Wilson pointed out that there can exist
an exact lattice version of chiral symmetry that will
likewise guarantee the continuum chiral symmetry
without fine-tuning.\cite{Ginsparg:1981bj}  There now exist realizations 
of the Ginsparg-Wilson relation that the lattice
Dirac operator must obey, introduced in the quite
famous papers of Kaplan, Neuberger\cite{Kaplan:1992bt} 
and L\"uscher.\cite{Luscher:1998pq}
In the latter paper L\"uscher suggested the possibility  
of an exact lattice supersymmetry.
This notion has been explicitly realized by several groups;
for example, see refs.~\ref{Kaplan:2002wv}-\ref{Catterall:2000rv} and 
references therein.  
In the examples that are interacting,
the supercharges are nilpotent, $Q^2=0$,
allowing for an algebra that is consistent with the absence of 
infinitesmal Poincar\'e invariance on the lattice.
The exact lattice supersymmetry forbids certain relevant or
marginal operators.
This eliminates or reduces the fine-tuning required to obtain
the target continuum theory.  One rather interesting idea
is based on deconstruction; it is due
to Cohen, Kaplan, Katz and \"Unsal.\cite{Kaplan:2002wv}${}^-$\cite{CKKU}  
They are able to write down lattice Yang-Mills theories with some
exact lattice supersymmetry.
Brief reviews by Kaplan are available; these
are readable and supply details that I do not have space to 
provide.\cite{Kaplan:2002zs}

\ite{Deconstruction, originally.}  This is a
gauge-invariant regularization of $d>4$ gauge theories.
One obtains an effective $d>4$ gauge theory
from a 4d {\it quiver} = {\it moose} = product group
gauge theory.\cite{Arkani-Hamed:2001ca}
1 or 2 dims are {\it emergent,} effective dimensions.
In Fig.~1, I show a quiver theory with 6 factors $G_1,\ldots,G_6$
and representations $({\bf R_i}, {\bf R_j})$ that are charged under pairs, forming
``links'' of a ``torus'' in ``theory space.''

\begin{figure}[p]
\begin{center}
\includegraphics[height=2.5in,width=2.5in]{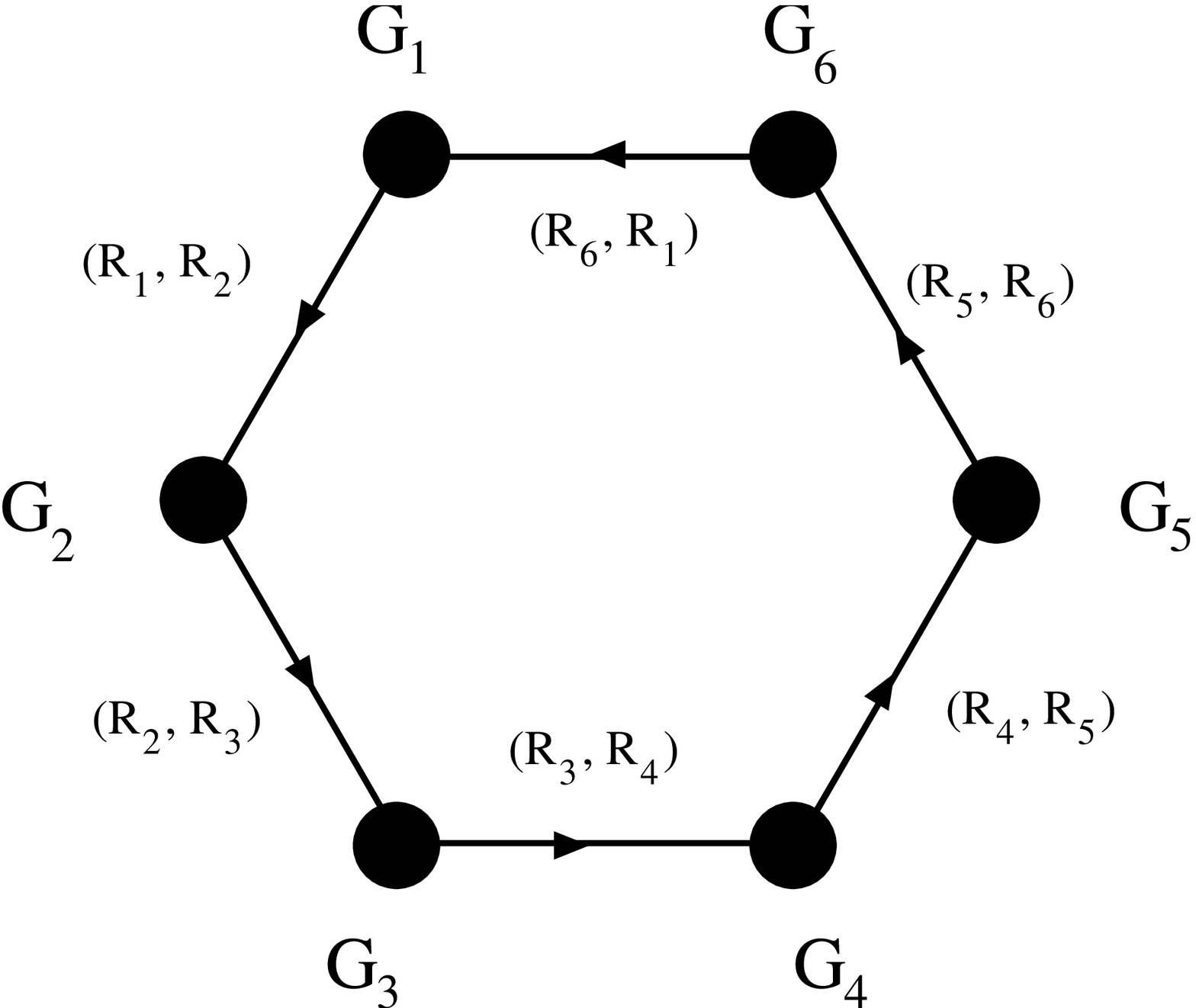}
\end{center}
\caption{Obligatory quiver diagram.}
\end{figure}

\ite{Deconstruction, latticization.}  Fig. 1 cries out:
Why not make some/all of our 4d emergent?
Is this a pathway to new lattice theories?
Maybe if the quiver is supersymmetric, will we
wind up with some exact lattice supersymmetry?

Deconstruction of all 3 spatial 
dimensions was considered by Kaplan, Katz and \"Unsal.\cite{Kaplan:2002wv}
Deconstruction of just 1 spatial dimension
was considered by Poppitz, Rozali and myself;\cite{Giedt:2003xr} 
we looked at the $U(1)_R$ anomaly in $\Ncal=2$ 4d SYM. 
Poppitz has gone on to analyze KK monopoles in this 
framework.\cite{Poppitz:2003uz}

The case with all dimensions latticized has also begun
to be studied.  Here one starts with 0d supersymmetric 
quiver theories.  This affords a spacetime (Euclidean) 
lattice formulation of SYM, and explicit constructions
have been written down by Cohen et al.\cite{Cohen:2003xe,CKKU}
These 0d theories contain some exact lattice supersymmetry
that eliminates or reduces fine-tuning.
I have studied some of the practical aspects of these 
theories;\cite{Giedt:2003ve,Giedt:2003vy}
i.e., relating to actually running a Monte Carlo simulation
that would compute Euclidean Green's functions.  I will
now outline a couple of the simpler models and comment very
briefly on my recent findings.

\ite{(2,2) 2d $U(k)$ SYM.}  This model was introduced in ref.~\ref{Cohen:2003xe}.
The target theory is (2,2) 2d $U(k)$ SYM,
which is obtained\footnote{All SYM theories
discussed in the following are the Euclidean versions.}
from a 4d $\to$ 2d reduction 
of $\Ncal=1$ 4d pure SYM.  We start with a {\it mother theory.}
It is the $\Ncal=4$ SYM matrix model
that is obtained from 4d $\to$ 0d reduction of 4d $\Ncal=1$ $U(kN^2)$
SYM.  The theory contains (Hermitian) bosons 
$v_m = v_m^\beta T^\beta, \; m \in \{ 0,1,2,3 \}$
in a Hermitian basis $T^\beta$ of $u(kN^2)$.
In addition the theory contains 2-component 
Euclidean Majorana-Nicolai fermions
$\psi = \psi^\beta T^\beta, \;
\psib = \psib^\beta T^\beta, \;
\psi = (\lambda, \xi)^T, \; \psib=(\alpha, \beta)$.
The mother theory action is
\beq
S = \frac{1}{4g^2} \tr \( [v_m, v_n][v_m, v_n] \)
+ \frac{1}{g^2}\tr \( \psib \sbar_m [v_m, \psi] \), \quad
\sbar_m = (1, i\sbf)
\eeq

Next we progress to a {\it daughter theory.}
We project out all fields {\it not} inert with respect to
a $Z_N \times Z_N$ symmetry of mother theory;
i.e., we orbifold the matrix model.
I suppress most of the details.
The orbifold is chosen in such a way that it
breaks the mother theory gauge group to a 
$U(k)$ lattice gauge symmetry:
$U(kN^2) \to \bigotimes_{\mbf=(1,1)}^{(N,N)} U(k)_\mbf$.
The bosons which survive decompose as follows:
$v_m \to x_\mbf, \xb_\mbf, y_\mbf, \yb_\mbf$,
with each a bifundamental linking two factors in
the quiver, just as in the 1d quiver mentioned above.
$x$'s are links in the $x$-direction and $y$'s
are links in the $y$-direction.

Finally we extract the {\it lattice theory.}
To do this we expand around a point in moduli space:
$ x_\mbf = (2 a^2)^{-1/2} \obf + \cdots, \;
y_\mbf = (2 a^2)^{-1/2} \obf + \cdots$.
This is stabilized with deformation
of the action:
\beq
\Delta S = \frac{a^2 \mu^2}{g^2} \sum_\mbf \tr \[
\(x_\mbf x_\mbf^\dagger - \frac{1}{2a^2}\)^2
+ \(y_\mbf y_\mbf^\dagger - \frac{1}{2a^2}\)^2 \]
\eeq
Although this breaks the lattice supersymmetry, we scale
the deformation away in the thermodynamic limit:
$\mu \sim 1/Na$.  Further conditions, required to control fluctuations,
are described in ref.~\ref{Cohen:2003xe}.

\ite{Fermion determinant.}  In the daughter theory,
the fermion action is of the form $S_F = \psib M \psi$,
and the (boson field dependent) matrix $M$ is problematic.\cite{Giedt:2003ve}
First, there exists an ever-present fermion zeromode,
$\det M \equiv 0$, independent of the boson configuration.
I project it out to exhibit the determinant for the other fermions.
A simple method is to deform the fermion matrix
$M \to M_\e \equiv  M + \e {\bf 1}$.
Then the product of nonzero eigenvalues is identical to
$\det \hat M_0 \equiv \lim_{\e \to 0^+} \e^{-1} \det M_\e$.
In the computation of Green's functions, this 
procedure factors out and cancels the zeromode.
I have studied $\det \hat M_0$ for boson configurations
drawn randomly from a Gaussian distribution,
centered on the relevant point in moduli space, with unit variance.
The phase $\phi = \arg \det \hat M_0$ is
uniformly distributed throughout the interval $(-\pi,\pi]$
(see Fig. 2).  

After integrating out the fermions, we are left with 
a complex ``effective'' boson action
\beq
\exp(-S_{\stxt{eff}}(x,y)) = e^{i \phi(x,y)} 
\[ \exp \( -S_B(x,y) + \ln |\det \hat M(0;x,y)| \) \]
\eeq
Presumably the corresponding violation of unitarity
vanishes in the continuum limit; certainly we see no
evidence of it in the classical continuum limit, which
has been worked out in ref.~\ref{Cohen:2003xe}.
The quantity in brackets is a valid probability measure
for boson configurations.  The corresponding action is called
the {\it phase-quenched} action:
$S_{p.q.}(x,y) = S_B(x,y) - \ln |\det \hat M(0;x,y)|$.
It is difficult to study the original action by Monte Carlo
methods, because we must reweight the phase-quenched expectation values
by the phase using the identity $\vev{{\cal O}} =
\vev{e^{i \phi}{\cal O}}_{p.q.}/\vev{e^{i \phi}}_{p.q.}$
The cancellations that occur as the phase flops around
are in most cases a severe problem.  The same problem is
encountered in QCD with an appreciable baryon density.

\begin{figure}
\begin{center}
\includegraphics[height=4in,width=2.5in,angle=90]{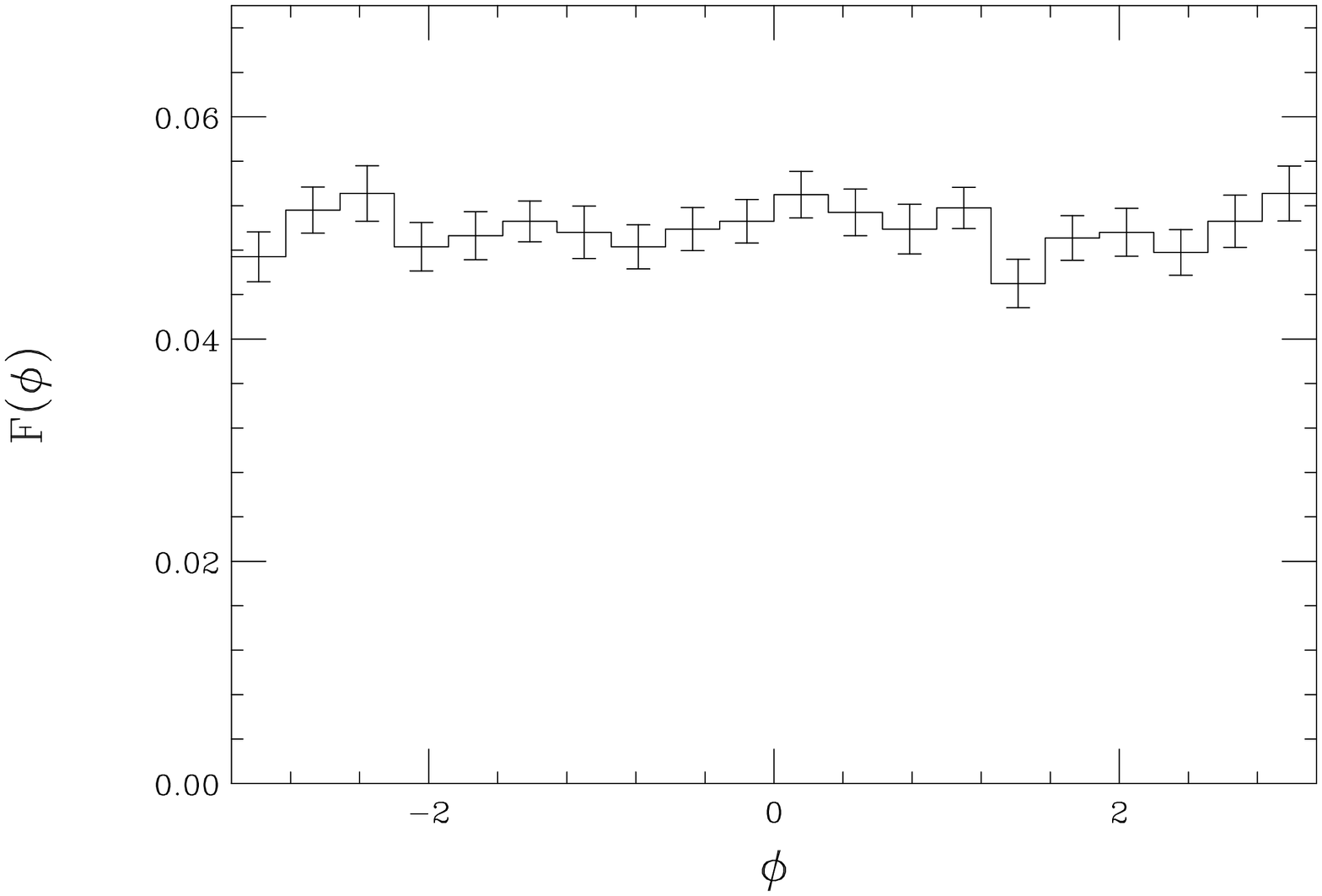}
\vskip 20pt
\caption{Frequency $F(\phi)$ of the phase
$\phi=\arg \det \hat M_0$ in all bins for a
Gaussian boson distribution on a $6 \times 6$ lattice
whose target is (2,2) 2d $U(2)$ SYM.
Smaller lattices give similar results.} 
\end{center}
\end{figure}

\ite{(4,4) 2d $U(k)$ SYM.}  This is another model constructed by
Cohen et al.\cite{CKKU}
The target theory  is (4,4) 2d $U(k)$ SYM,
obtained from the 6d $\to$ 2d reduction of $\Ncal=1$ 6d pure $U(k)$ SYM.
For this model I find similar problems fermion determinant.\cite{Giedt:2003vy}
With random boson pulls from a Gaussian distribution
I obtain a roughly flat distribution for the phase
of the fermion determinant; the data is quite similar
to Fig. 2.  One might hope
that in the phase-quenched ensemble the situation
would improve---that the phase of the fermion determinant
would most often be found near some particular value.
Unfortunately, I find that this is not the case for the smaller
lattices that I can presently simulate; again, the data is flat like Fig. 2.

\ite{Conclusions.}  While it would be very nice indeed to obtain
reliable and accurate results for supersymmetric Yang-Mills using
lattice simulations, it is not at all an easy task.  Further
research is needed; it remains to be seen whether or not
results useful to string phenomenology will emerge.

\vspace{15pt}

\noindent
{\bf Acknowledgements} \vspace{5pt} \\
I thank Erich Poppitz for collaboration and fruitful conversation.
I received support from the National Science and Engineering 
Research Council of Canada and from the Ontario 
Premier's Research Excellence Award.

\end{document}